\documentclass[twocolumn]{cinc}
\usepackage{graphicx}
\usepackage{hyperref}
\usepackage{mathtools}
\usepackage{amsmath,amssymb,latexsym}
\usepackage{booktabs}
\usepackage{scrextend}
\hypersetup{
    colorlinks=true,
    linkcolor=black,
    filecolor=magenta,      
    urlcolor=blue,
    citecolor = black
}
\deffootnote[1em]{1em}{1em}{\textsuperscript{\thefootnotemark}}
\begin{document}
\bibliographystyle{cinc}

\title{Generative Pre-Trained Transformer for Cardiac Abnormality Detection}

\author {Pierre Louis Gaudilliere$^{1}$, Halla Sigurthorsdottir$^{1}$, Clémentine Aguet$^{1}$, \\ Jérôme Van Zaen$^{1}$, Mathieu Lemay$^{1}$, Ricard Delgado-Gonzalo$^{1}$ \\
\ \\ 
 $^1$ Swiss Center for Electronics and Microtechnology (CSEM SA), Neuchâtel, Switzerland }

\maketitle

\begin{abstract}

ECG heartbeat classification plays a vital role in diagnosis of cardiac arrhythmia. The goal of the Physionet/CinC 2021 challenge was to accurately classify clinical diagnosis based on 12, 6, 4, 3 or 2-lead ECG recordings in order to aid doctors in the diagnoses of different heart conditions.  Transformers have had great success in the field of natural language processing in the past years. Our team, CinCSEM, proposes to draw the parallel between text and periodic time series signals by viewing the repeated period as “words” and the whole signal as a sequence of such words. In this way, the attention mechanisms of the transformers can be applied to periodic time series signals. In our implementation, we follow the Transformer Encoder architecture, which combines several encoder layers followed by a dense layer with linear or sigmoid activation for generative pre-training or classification, respectively. The use case presented here is multi-label classification of heartbeat abnormalities of ECG recordings shared by the challenge. Our best entry, not exceeding the challenge's hardware limitations, achieved a score of 0.12, 0.07, 0.10, 0.10 and 0.07 on 12-lead, 6-lead, 4-lead, 3-lead and 2-lead test set respectively. Unfortunately, our team was unable to be ranked because of a missing pre-print.

\end{abstract}

\section{Introduction}

In the recent years transformers~\cite{vaswani} have shown to perform strongly in the field of natural language processing. By solving the problem of long-term dependencies and effectively paying attention to the right words for the context, very impressive transformer-based language generation and translation models have been created including Google’s BERT~\cite{devlin} and OpenAI’s GPT-3~\cite{brown}.

A sentence, or a piece of text, is a series of words that represents a finite amount of repeated patterns with a semantic meaning. Similarly, certain time series signals such as electrocardiogram (ECG signals) can be thought of as series of a finite amount of repeated patterns (heartbeats). Here we exploit this similarity by modeling single heartbeats as words and full ECG signals as sentences.

The contribution of the present work is two-fold:

\begin{itemize}
	\item Drawing the conceptual parallel between periodic time series signals and text
	\item Using that to create a simple, yet powerful, representation of ECG signals, representing each heartbeat as a “word embedding” and an ECG signal as a sequence of such “words”\\
\end{itemize}

The paper is organized as follows. Section~\ref{Previous Work} exposes the background. In Section~\ref{Methods}, we describe the pipeline from raw ECG signals to multi-label classification, as well as the model architecture. Then, in Section~\ref{Results}, we expose the obtained results. Finally, in Section~\ref{Discussion and Conclusions}, we give a critical review of the implementation and propose further lines of research.

\section{Previous Work}\label{Previous Work}

In the recent years, transformers have been used on time-series based tasks~\cite{song, wu}. Regarding ECG classification, Yan et al.~\cite{yan} got good results on the MIT-BIH database, i.e. a per heartbeat classification task. Furthermore,~\cite{natarajan} used a convolutional neural network (CNN)+transformer architecture alongside hand-crafted features to win the CinC2020 challenge~\cite{perez}, where each recording (duration between 5~s and 30~min) was weakly labelled (a multi-label classification problem).
However, none of these works drew the parallel between their time-series signals and text. Furthermore, none of the ECG-based works used a heartbeat as a word embedding, nor used their transformer on a sequence of periodic “words”.

\begin{figure*}[htbp]
\centering
\includegraphics[width=\textwidth]{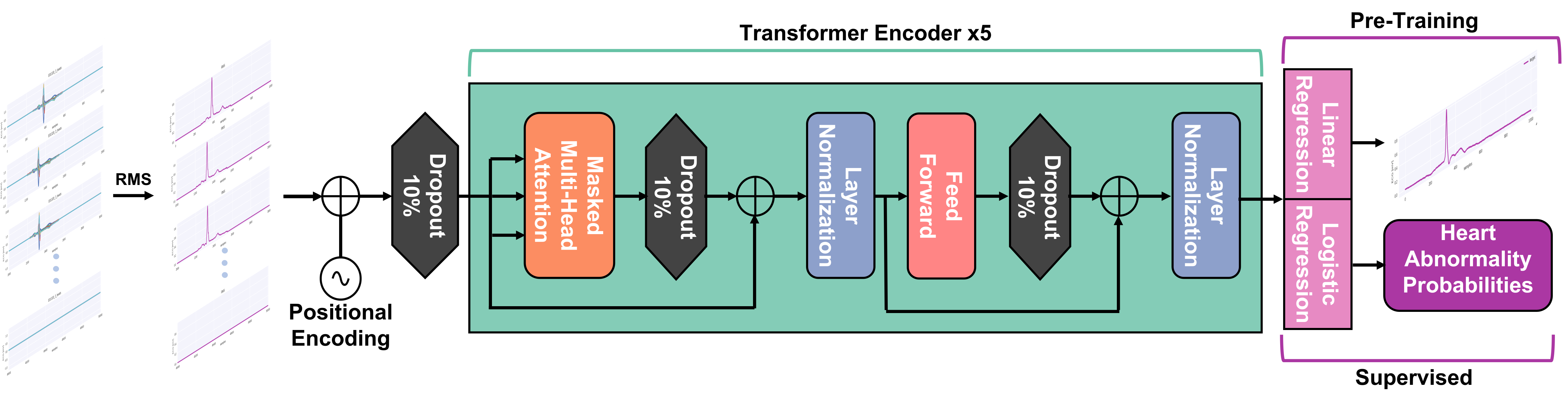}
\caption{Overview of the complete system: we split the raw ECG signal into a sequence of heartbeats that is fed into a transformer. The last fully connected layer takes the output of the transformer to produce a prediction of the next heartbeat or to output probabilities for multi-label classification of cardiac arrhythmia.}
\label{transformer}
\end{figure*}

\section{Methods}\label{Methods}

In this work, we present a deep transformer network designed for multi-label classification of $d_{classes} = 28$ cardiac arrhythmia including atrial fibrillation, atrial flutter or premature atrial contraction. Our network uses sequences of heartbeats embedded as “words”. The latter are fed into a Transformer architecture~\cite{vaswani} that relies entirely on a parallelizable self-attention mechanism. We first pre-trained the model so that it learns about features of the ECG waveforms and to make the model more generalizable by adding external data sources\footnote{\url{https://physionet.org/content/edb/1.0.0/}}\footnote{\url{https://physionet.org/content/mitdb/1.0.0/}}. During pre-training, the model receives as input a sequence of single heartbeats and it outputs a guess of the next heartbeat. Therefore, we fed the output of the transformer into a fully connected layer with $d_{model} = 1000$ output neurons and linear activation. Finally, for the classification task (supervised learning), we used transfer learning and replaced the last layer with a fully connected layer with 28 output probabilities and sigmoid activation. Figure~\ref{transformer} gives an overview of the complete system.

\subsection{Dataset} 
 
The database combines eight worldwide datasets, for a total of 88k ECG recordings shared for training~\cite{cinc2021}. we left aside recordings from St Petersburg since they are 30~min long and more susceptible to artifacts such as movement. Moreover, we restricted the total number of heartbeats in a sequence to be $maximum\_position\_encoding = 50$. We truncated longer sequences, whereas we zero padded shorter sequences at the end for the missing heartbeats.  We separated the remaining recordings into five folds, stratified by labels and datasets. We assigned four of the folds to training while we preserved the fifth for validation. To cope for possible inconsistencies in labeling between datasets, we trained on different combination of datasets. All the challenge datasets were included for validation.
Some labels (SA, TInv) are underrepresented which make it harder for the model to learn a good representation. We combined the labels that were considered equivalent by the challenge. Finally, we did not assign a label to recordings that have only non-scored labels.

\subsection{Signal Pre-Processing}  

To be able to properly view the signal as a series of repeated heartbeats, we need to perform some pre-processing steps. First, we scaled the ECG leads by their respective ADC gain. Then, we applied an IIR (infinite impulse response) high-pass filter with a cutoff frequency of 0.5~Hz. We then tested four R-peak detectors\footnote{\url{https://github.com/berndporr/py-ecg-detectors}} (Christov, Hamilton, Two-Average and Pan Tompkins detector) in order to separate the raw ECG signals into heartbeats. Finally, to uniformize the sampling frequency between the datasets, we re-sampled each recording to 500~Hz, and we updated the timing of the R-peaks accordingly.

\subsection{Word Embeddings} 

\begin{figure}[htbp]
\includegraphics[width=\linewidth]{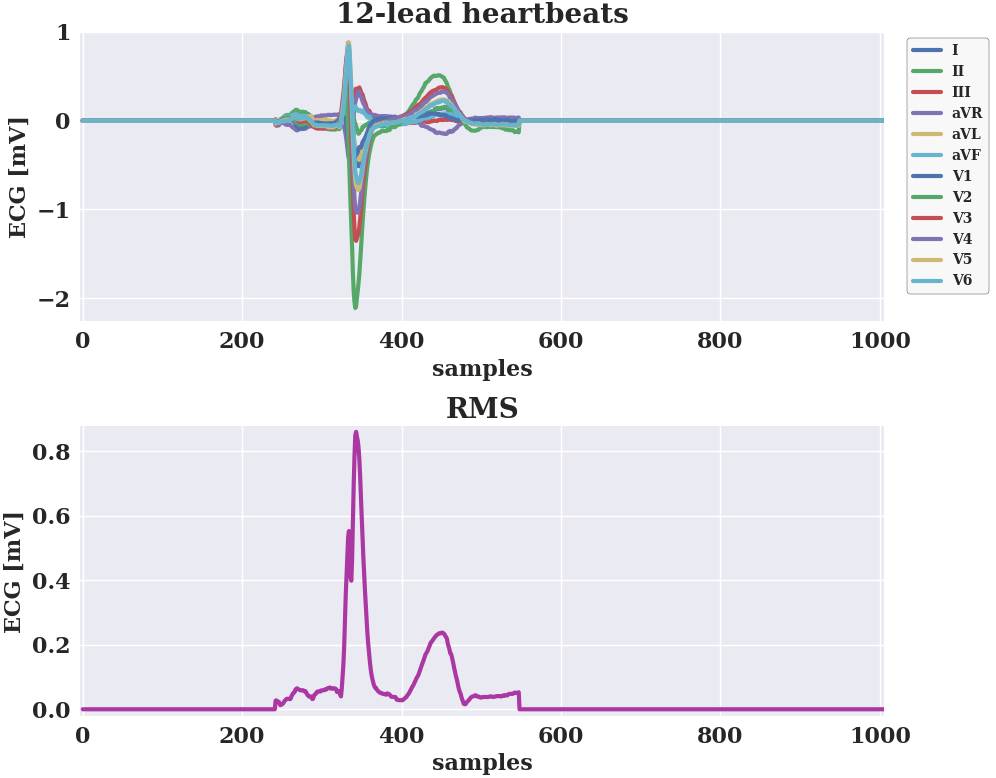}
\caption{We aligned the heartbeats at the R-peaks and we zero padded on the left and right for a total length of 1000 samples. Then, we computed the RMS of the active leads.}
\label{word_embedding}
\end{figure}

The first step in a transformer model is to create word embeddings. These embeddings represent the word in an embedding space. Here, we tested the most simplistic way of representing a heartbeat as a “word embedding”. We considered 1/3 of the R-R interval before the R-peak and 2/3 of the R-R interval after the R-peak to be a part of the heartbeat. We then aligned the heartbeats at the R-peaks. To create embeddings of the same length, we defined a maximum normal length of a heartbeat to be $d_{model} = 1000$ samples, or 2~s. We zero padded any heartbeat below this length and we truncated any heartbeat above this length. Therefore, one can infer the variability of the R-R interval by looking at the number of added zeros: the shorter the interval the more we padded with zeros. Finally, we computed the root mean square (RMS) of the active ECG leads to obtain a single ECG signal (Figure~\ref{word_embedding}).

\subsection{Transformer Model}

As transformers rely uniquely on self-attention, we use positional encoding~\cite{vaswani} in order to infer the relative timing of the heartbeats within a sequence. We added together the embedding representations $(x_0, ... , x_n)$, where  $x_i \in R^{d_{model}}$, with the positional encodings $(p_0, ..., p_n)$ and the result is given as input to the transformer.

Our transformer uses a stack of $N = 5$ encoders. The input of each encoder is a masked multi-head attention layer followed by a feed forward network which combines two dense layers with $\mathit{dff}$ and $d_{model}$ output neurons respectively. For regularization, we applied a dropout at the output of each layer before it is added to the input of the layer and normalized. 

The masked multi-head attention layer at the beginning of the transformer is composed of $h = 8$ heads. Each embedding representation $x_i$ is passed to a dense linear layer of $d_{model}$ units to create a query, key and value vector ($q, k,$ and $v$ respectively). These vectors are split equally between the heads and stacked into matrices $\bold{Q}, \bold{K}$ and $\bold{V}$, whose depth is then equal to $d_{q,k,v} = d_{model} // h = 125$. Finally, self-attention is computed using equation~(\ref{attention_equation}).
\begin{gather}
	Attention(\bold{Q}, \bold{K}, \bold{V}) = softmax\left(\frac{\bold{QK}^T}{\sqrt{d_k}}\right)\bold{V}
\label{attention_equation}
\end{gather}

The output of the transformer is fed into a last fully connected layer either with $d_{model}$ output neurons and linear activation, or with $d_{class}$ output probabilities and sigmoid activation.

\subsection{Training}

We deemed a class positive if the output probability of the last layer for that class is greater than a fixed $threshold = 0.5$. When pre-training the model, the loss function is the mean squared error whereas during supervised training, the loss function is the standard binary cross entropy loss averaged across the classes. During pre-training, for a sequence of $n$ heartbeats, we took the first $i$ ($1\leq i\leq n-1$) heartbeats and the model had to generate a prediction of the next $i+1$ heartbeat. We used the Adam optimizer~\cite{vaswani} ($\beta_1 = 0.9, \beta_2 = 0.98$ and $\epsilon = 10^{-9}$) whose learning rate varies following equation~(\ref{learning_rate}), with $warmup\_steps = 4000$.
\begin{gather}
	lr = \frac{1}{\sqrt{d_{model}}}\cdot \min \left(\frac{1}{\sqrt{step\_num}},\frac{step\_num}{warmup\_steps^{1.5}}\right)
\label{learning_rate}
\end{gather}

The complete model is composed of $90,536,240$ trainable parameters, initialized using Xavier uniform initialization. we trained the models over 50 epochs on subsets of the data made publicly available for the 2021 Physionet/CinC challenge.  We ran the different models using Tensorflow:2.5.0-gpu on three RTX 2080 Ti Turbo GPUs. We selected the hyperparameters described in Table~\ref{hyperparameters} from the references.

\newcommand{\ra}[1]{\renewcommand{\arraystretch}{#1}}
\begin{table}[htbp]

\caption{\label{hyperparameters} List of selected hyperparameters to train the model}
\vspace{4 mm}
\ra{1.3}
\centerline{\begin{tabular}{@{}l|r@{}} 
\toprule[1.2pt]
Hyperparameters    & Value\\ \midrule[1.2pt]
Sampling Frequency~Hz & 500 \\
Batch size     & 128 \\
Number of classes, $d_{class}$   & 28 \\
Threshold & 0.5 \\
Dropout & 0.1 \\
Number of encoders, $N$ & 5 \\
Maximum position encoding & 50\\
Embedding size, $d_{model}$  & 1000 \\
Number of heads, $h$   & 8 \\
Feed forward layer, $\mathit{dff}$ &  2048 \\
Depth of query, key, and value vectors, $d_{q,k,v}$  & 125\\ 
\bottomrule[1.2pt]
\end{tabular}}

\end{table}

\begin{table*}
\caption{\label{model_score} CinC 2021 validation and final test scores and running times}
\vspace{4 mm}
\ra{1.3}
\centerline{\begin{tabular}{@{}lccccccrrr@{}} 
\toprule[1.2pt]
Leads & \multicolumn{6}{c}{Challenge Datasets} & \multicolumn{3}{r}{Running Time (minutes)}\\ 
\cmidrule[0.5pt]{2-10}
& Validation set & CPSC test & G12EC test & Undisclosed test & UMich test & Test set & Train & Validation & Test \\
\midrule[1.2pt]
All-lead & 0.10 & 0.15 & 0.10 & 0.08 & 0.10 & 0.10 & 677 & 29 & 115\\ 
12-lead & 0.13 & 0.20 & 0.12 & 0.09 & 0.13 & 0.12 & 677 & 29 & 115\\
6-lead & 0.07 & 0.04 & 0.08 & 0.09 & 0.05 & 0.07 & 677 & 26 & 112\\
4-lead & 0.12 & 0.18 & 0.11 & 0.08 & 0.11 & 0.10 & 677 & 26 & 120\\
3-lead & 0.11 & 0.18 & 0.10 & 0.07 & 0.10 & 0.10 & 677 & 28 & 111\\
2-lead & 0.07 & 0.07 & 0.08 & 0.07 & 0.07 & 0.07 & 677 & 26 & 112\\
\bottomrule[1.2pt]
\end{tabular}}

\end{table*}

\section{Results}\label{Results}

We pre-trained the model on external data sources as well as St Petersburg. However, because of their length (30~min long), the data was often corrupted and lead to poor prediction. After 20 epochs the loss converged to $2.99\mathrm{E}{-5}$. We then decided to pre-train the model on Georgia and PTB\_XL and the loss converged to $1.90\mathrm{E}{-5}$ after 5 epochs. However, neither linear probe nor transfer learning of the pre-trained model improved the score as opposed to supervised learning alone. Therefore, pre-training was not included in the final model.

We first trained the model on Georgia and PTB\_XL. Then, adding Chapman and Ningbo data sources positively impacted the performance. However, when training on the full database, our model exceeded the hardware limitation. 

The best model utilized the Two-Average R-peak detector and was trained on Georgia, PTB\_XL, Chapman and Ningbo. Unfortunately, we did not manage to obtain a successful entry with that model. Our best entry, the model trained on Geogia and PTB\_XL with the Christov detector, achieved a score of 0.12, 0.07, 0.10, 0.10 and 0.07 on 12-lead, 6-lead, 4-lead, 3-lead and 2-lead test set respectively (Table~\ref{model_score}). 

\section{Discussion and Conclusions}\label{Discussion and Conclusions}

Although the use of transformers and the interpretation of ECG signal as sentences and heartbeats as words seem promising, our model did not manage to obtain a satisfying score. Neither linear probe nor transfer learning of the pre-trained model improved the score as opposed to supervised training alone. Our approach is highly dependent on the R-peak detectors used. Open-source R-peak detectors are not perfect and wrong R-peak detection could be a major source of error. Finally, although the RMS provides good generalization, it is interesting to explore other word embedding methods such as the CNN developed by Sigurthorsdottir et al. 2020~\cite{halla}.

\bibliography{refs}

\begin{correspondence}
Pierre Louis Gaudilliere\\
Rue Jaquet-Droz 1, Neuchâtel, NE, Switzerland\\
pierre.gaudilliere@csem.ch
\end{correspondence}

\end{document}